\documentclass[aps,prl,twocolumn,superscriptaddress]{revtex4}

\usepackage{graphicx}
\usepackage[german,english]{babel}
\newcommand{\diff}{\ensuremath{\text{d}}}

\newcommand{\imp}{\rm{imp}}

\renewcommand{\Re}{{\rm Re}}
\renewcommand{\Im}{{\rm Im}}

\newcommand{\beqa}{\begin{eqnarray}}
\newcommand{\eeqa}{\end{eqnarray}}

\begin{document}

\title{Impurity States in Graphene}

\author{T. O. Wehling}
\affiliation{I. Institut f{\"u}r Theoretische Physik, Universit{\"a}t Hamburg, Jungiusstra{\ss}e 9, D-20355 Hamburg, Germany}
\author{A. V.  Balatsky$^{*}$}
\affiliation{Theoretical Division, Los Alamos National Laboratory,
Los Alamos, New Mexico 87545,USA}
\email[]{avb@lanl.gov, http://theory.lanl.gov}
\author{M. I. Katsnelson}
\affiliation{Institute for Molecules and Materials,University of Nijmegen, Toernooiveld 1, 6525 ED Nijmegen, The Netherlands}
\author{A. I. Lichtenstein}
\affiliation{I. Institut f{\"u}r Theoretische Physik, Universit{\"a}t Hamburg, Jungiusstra{\ss}e 9, D-20355 Hamburg, Germany}
\author{K. Scharnberg}
\affiliation{I. Institut f{\"u}r Theoretische Physik, Universit{\"a}t Hamburg, Jungiusstra{\ss}e 9, D-20355 Hamburg, Germany}
\author{R. Wiesendanger}
\affiliation{I. Institut f{\"u}r Angewandte Physik, Universit{\"a}t Hamburg, Jungiusstra{\ss}e 11, D-20355 Hamburg, Germany}


\date{\today}

\begin{abstract}
Defects in graphene are of crucial importance for its electronic and
magnetic properties. Here impurity effects on the electronic
structure of surrounding carbon atoms are considered and the
distribution of the local densities of states (LDOS) is calculated. As
the full range from near field to the asymptotic regime is covered,
our results are directly accessible by scanning tunnelling
microscopy (STM). We also include exchange scattering at magnetic
impurities and eludicate how strongly spin polarized impurity states
arise.
\end{abstract}



\maketitle

Graphene, a recently discovered allotrope of carbon and the first known example of a truly
two-dimensional (2D) crystal \cite{Novoselov_science2004,K.S.Novoselov07262005} has
unique electronic properties \cite{Geim2005, Zhang2005, vozmediano:155121, peres:174406, pereira2006, pogorelov-2006-, peres:125411, Loktev-2006-, Katsnelson:zitter, morozov:016801, cheianov-2006-, cheianov-2006b-, Novoselov_NPhys2006}, such as an exotic quantum
Hall effect with half-integer quantization of the Hall conductivity
\cite{Geim2005, Zhang2005}, finite conductivity at zero
charge-carrier concentration \cite{Geim2005}, strong suppression of
weak localization \cite{morozov:016801}, etc. The peculiar 2D band
structure of graphene resembles ultrarelativistic electron dynamics
near two nodal points in the Brillouin zone. This provides a new
bridge between condensed matter theory and quantum electrodynamics
(index theorem and the half-integer quantum Hall effect
\cite{Geim2005}, relativistic Zitterbewegung
\cite{Zitterbewegung} and the minimal
conductivity \cite{Katsnelson:zitter}, ``Klein paradox''
\cite{KleinParadox} and anomalous tunnelling of electrons in graphene through potential
barriers \cite{Novoselov_NPhys2006}). Unexpectedly high electron
mobility in graphene and its perfect suitability for planar
technology makes it a perspective material for a next-generation,
carbon-based electronics \cite{Novoselov_science2004}.

Impurity states are important contributors to these unusual properties. Graphene is conducting due to carriers that can
either be introduced by a gate voltage \cite{Novoselov_science2004, Zhang2005, Geim2005} or by doping  \cite{Novoselov_science2004, peres:125411, pogorelov-2006-}. This situation is very reminiscent of doped semiconductors, where the desired properties are obtained by creating an impurity band. Recent progress in scanning tunneling microscopy (STM) made it possible to image impurity states for a wide class of materials with very high spatial resolution. This so-called ``wave function imaging'' yields local images of the impurity-induced wave function. Examples of wave function imaging range from unconventional superconductors \cite{Pan2000, Hudson2001} to semiconductors \cite{maltezopoulos:196804, Yazdani2006}, magnetic metals \cite{bergmann:2004} and graphite surfaces \cite{MIZES1989, Kelly1996, Kelly2000}. It allows one to investigate the formation of the impurity band and the associated electronic properties. Theoretical modelling and STM measurements of near impurity site effects can be compared and thus eludicate, e.g., magnetic interaction mechanisms \cite{Yazdani2006, Tang2004}.

The purpose of this letter is to address the question of electronic
properties of single and double impurities in graphene in connection with future STM
experiments and impurity-induced ferromagnetism. The impurity states
are characterized by their energy and by their real space wave
functions that determine the shape of the resonance. In contrast to
previous studies \cite{peres:125411, cheianov-2006b-} we
consider the real space structure of the electronic state in the range from the
impurity site to the asymptotic regime, its dependence on the potential strength and the
spin exchange interaction. We will address the 
impurity states as an input into a subsequent discussion of
impurity-induced bands.

The honeycomb arrangement of carbon atoms in graphene can be described by a hexagonal
lattice with two sublattices A and B. (See, e.g., \cite{Semenoff-1984}).
 With the Fermi operators $c_i$ and $d_i$ of
 electrons in cell $i$ at sublattice A and B, respectively, we describe a single and two neighbouring
 impurities
  by $V_{\rm s}=U_0c^\dagger_0c_0$ and $V_{\rm d}=U_0(c^\dagger_0c_0+d^\dagger_0d_0)+U_1(c^\dagger_0
   d_0+d^\dagger_0
   c_0)$. Here $U_0$ is the potential strength and $U_1$ the change of sublattice hopping between
   the two impurity sites. Related to the current research are questions about impurities in graphite 
   that  have  been studied with STM \cite{MIZES1989}.  Only the atoms above
hollow sites are seen in STM on graphite. We find that
impurity states in graphene are {\em qualitatively} different from those in graphite
because of the sublattice degeneracy that is reflected in a
complicated sublattice structure of impurity induced resonances.

We find that impurity scattering produces low energy resonances with the real space
 structure and the resonant energy $E_{\imp}$ as function of $U_0$ (and $U_1$)
 clearly distinguishing between single and double impurities. For single impurities we
  find in agreement with Loktev \cite{Loktev-2006-}, that $E_{\imp}$ is well described by \beqa\label{eqn:eimp_U0_single} U_0=\frac{W^2}{E_{\imp}\ln\left|\frac{E{\imp}^2}{W^2-E{\imp}^2}\right|},\eeqa where W is the band
width. Hence the resonance energy $E_{\imp}$ approaches zero for $U_0\rightarrow\infty$.
Only strong single impurities (i.e. $U_0\gtrsim 10$\,eV) are capable of producing resonances
 within $1$\,eV of the Dirac point. This result is similar to the impurity states
observed in unconventional superconductors with Dirac
spectrum \cite{balatsky:373}.

The resonance of a double impurity is basically determined by $U_0-U_1$. Its energy coincides with the
 Dirac point at \textit{finite} $U_0-U_1=3t$, where $t\approx 2.7$\,eV is the nearest
 neighbour hopping parameter of graphene.

We give a detailed description of the local density of states (LDOS), the real space fingerprint of impurities
in graphene, both near the impurity and in the asymptotic regime at large distances. Near the impurity site that LDOS exhibits an intricate pattern. A single strong impurity placed on one sublattice produces a peak in LDOS at low energies that is large on the other sublattice. At large distances these impurity resonances have wave functions $\psi$ that asymptotically decay as $|\psi|^2\propto 1/r$.

We will consider potential scattering (nonmagnetic) as well as a magnetic impurities, i.e. spin dependent scattering. In the latter case the impurity induced resonance will exhibit a spin dependent splitting that might lead to a strong spin polarization of the impurity state. This observation, we believe, is important for the the discussion of moment formation and possible magnetic order in graphene.

To start with our \textit{theoretical model}, we describe the carbon ${\rm p_z}$-electrons within the tight binding approximation by $H  = \int_{\Omega_{\rm B}}\frac{\diff^2k}{\Omega_{\rm B}} \Psi^\dagger(k)H_k\Psi(k)$ with $\Psi(k)=\left[\begin{array}{c}c(k) \\ d(k)\end{array}\right]$ and $H_k  =  \left( \begin{array}{cc}
0 & \xi(k)\\
\xi^\ast(k)& 0 \end{array} \right)$,
where $\xi(k)=t\sum_{j=1}^3e^{ik(b_j-b_1)}.$
$\Omega_{\rm B}$ denotes the Brillouin zone volume and $c(k)$ ($d(k)$) are the $k$-space counterparts of $c_i$ ($d_i$).\cite{Semenoff-1984}

\begin{figure}
\centering
\includegraphics[width=.75\linewidth]{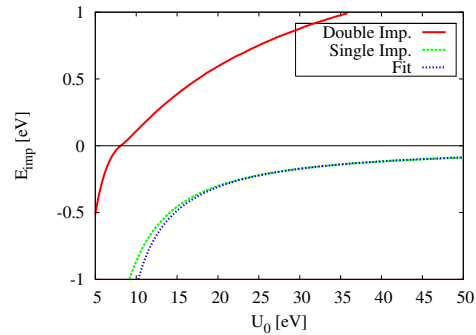}
\caption{\label{fig:impurity_energy}(Color online) The energy $E_{\imp}$ of the impurity resonance as function of the potential strength $U_0$ is shown for single impurities and double impurities with $U_1=0$. For the single scatterer $E_{\imp}$ obtained from our tight binding calculation is compared to the result obtained from the fully linearized bandstructure (Eqn. \ref{eqn:eimp_U0_single}) with fitted bandwidth $W=6.06$\,eV.}
\end{figure}

The full Green's function in real space $G(i,j,E)$ will be obtained using the $T$ matrix formalism: \begin{equation}
G(i,j,E)=G^0(i-j,E)+G^0(i,E)T(E)G^0(-j,E)\label{eqn:Gfull}.
\end{equation}
Therefore the unperturbed Green's function $G^0(i,E)$ in real space is calculated from its k-space counterpart $\tilde{G}^0(k,E)=(E-H_k+i\delta_E)^{-1}$ by Fourier transformation. Numerical problems in carrying out the Fourier integrals are avoided, by linearizing the bandstructure in a vicinity of the Dirac points, where all singularities occur. Outside these regions the full tight-binding bandstructure is taken into account.
Finally the $T$ matrix is given by
\begin{equation}
 T(E)=\left(\mathbf{1}-\tilde{V}_{\rm s(d)}G^0(0,E)\right)^{-1}\tilde{V}_{\rm s(d)} \label{eqn:T-matrix}
\end{equation}
with $\tilde{V}_{\rm s}=U_0\left( \begin{array}{cc}
1 &0\\
0 &0\end{array} \right)
$
and
$\tilde{V}_{\rm d}=\left( \begin{array}{cc}
U_0 &U_1\\
U_1 &U_0\end{array} \right)$ being the impurity potentials in k-space and matrix form.
Poles of the $T$-matrix  corresponding to impurity resonances occur, if
$\det(\mathbf{1}-\tilde{V}_{\rm s,d}G^0(0,E))=0$, i.e
\begin{equation}
\label{eqn:Eimp_single}
U_0G^0_{11}(0,E)-1=0
\end{equation}
for a single scatterer and
\begin{equation}
\label{eqn:Eimp_scalar}
(1-U_0G^0_{11}(0,E))^2-U_0^2G^0_{21}(0,E)G^0_{12}(0,E)=0
\end{equation}
for double impurity with $U_1=0$ --- this case, we refer to as scalar impurity. Near the Dirac points we have $|\Re(G^0)|\gg|\Im(G^0)|$, so that the impurity resonances $E_{\imp}$ as function of $U_0$ can be calculated from the previous two equations considering only the $\Re(G^0)$: The resulting real \textit{impurity energies} as a function of $U_0$ are shown in figure \ref{fig:impurity_energy}.
Adjusting the bandwidth parameter $W$ in Eqn. (\ref{eqn:eimp_U0_single}) to fit our $E_{\imp}(U_0)$ yields $W=6.06\pm 0.02$\,eV. I.e. $W$ differs slightly from the bandwidth $\overline{W}=\hbar v_{\rm f} \sqrt{\frac{\Omega_{\rm B}}{2\pi}}\approx 6.3$\,eV obtained by assuming linear dispersion in the entire Brillouin zone. \cite{Loktev-2006-}

It is quite remarkable that a pair of neighbouring scatterers produces a resonance at the Dirac point for $U_0=3t\approx 8.1$\,eV, while for a single impurity this occurs only in the limit of infinite potential strength. This effect can be attributed to the existence of \textit{two} nonequivalent Dirac points in the Brillouin zone. As a consequence at $E=0$ the onsite Green's function $G^0(0,0)$ has finite off diagonal $G^0_{12}(0,0)=G^0_{21}(0,0)=-\frac{1}{3t}$ but vanishing diagonal components resulting via equations \ref{eqn:Eimp_single} and \ref{eqn:Eimp_scalar} in the characteristic $E_{\imp}(U_0)$ curves.

For double impurities with sublattice hopping change $U_1$, it
follows directly from the secular equation, that the impurity energy
as a function of $U_0$ and $U_1$ is obtained from the scalar case by
replacing $U_0$ with $U_0-U_1$.

\begin{figure}
\begin{minipage}{.45\linewidth}
\centering
\includegraphics[width=.99\linewidth]{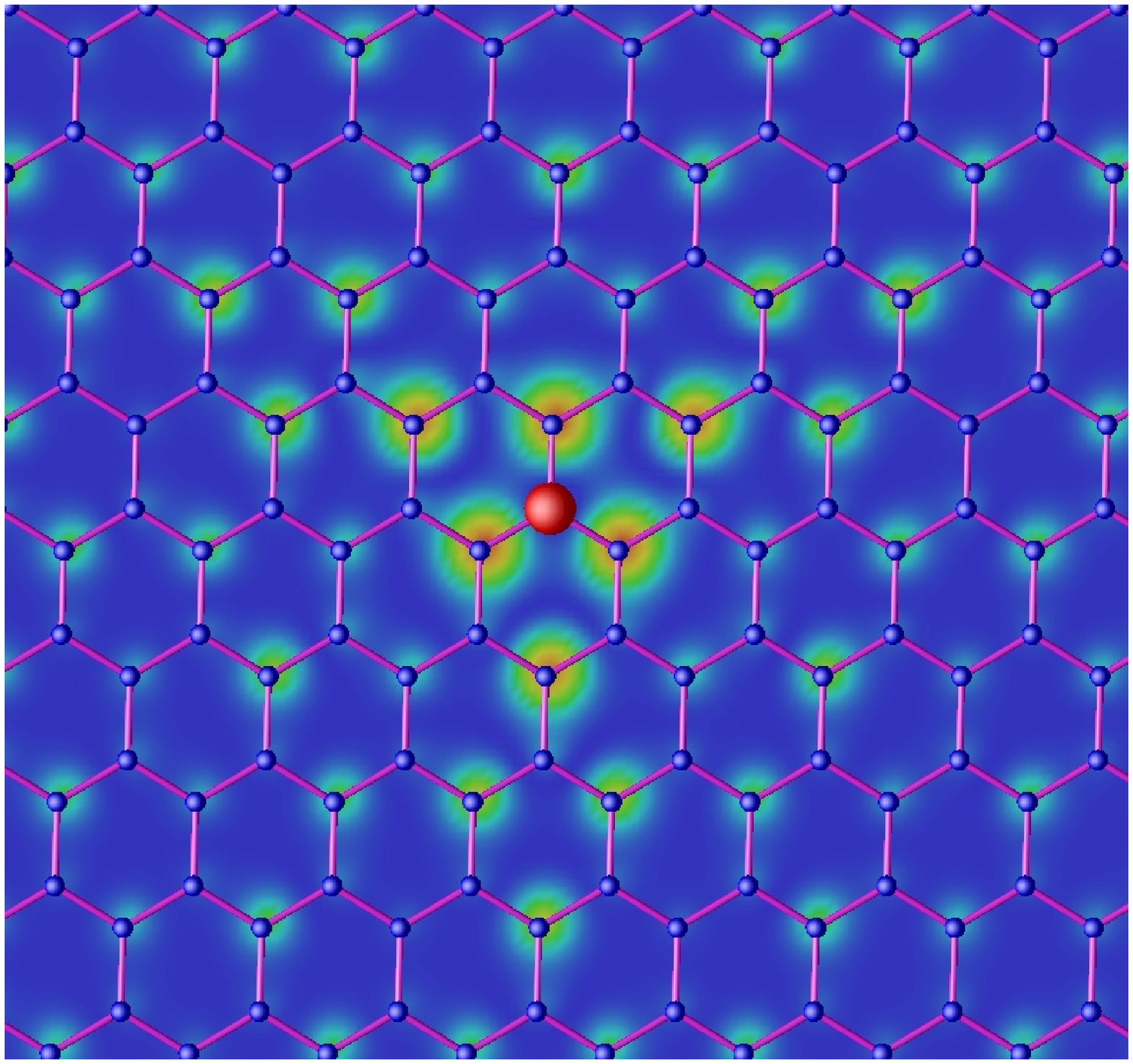}
\end{minipage}\hfill\begin{minipage}{.45\linewidth}
\centering
\includegraphics[width=.99\linewidth]{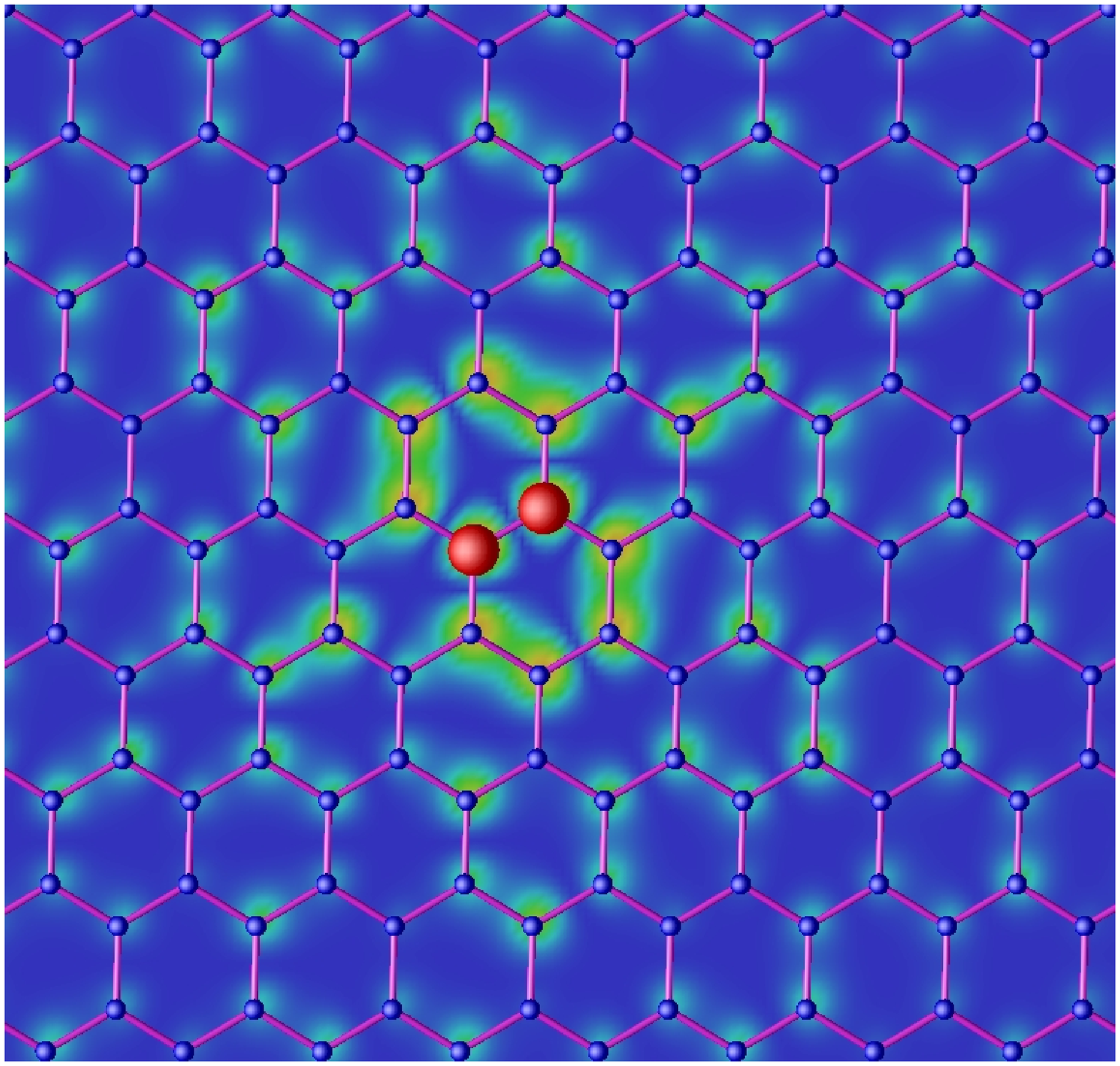}
\end{minipage}\begin{minipage}{.07\linewidth}
\centering
\includegraphics[width=.99\linewidth]{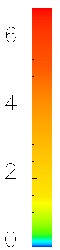}
\end{minipage}
\caption{\label{fig:NallE01x6}(Color online) $r$-dependent LDOS at $E=E_{\imp}=-0.1$\,eV for a single
impurity with $U_0=45$\,eV (left)  and for a scalar double impurity with $U_0=6.9$\,eV (right) encoded corresponding to color bar. The impurity sites are marked as big red dots in the center of the images.}
\end{figure}

We obtained the LDOS $N(r,E) = -\frac{1}{\pi}\Im\left(\sum_{i,j} \Phi_i(r) G(i,j,E)\Phi^\dagger_j(r)\right)$ in presence of impurities as a function of position and energy, where $\Phi_i(r)=\left[\phi_{i}^c(r),\phi_{i}^d(r)\right]$ with $\phi_{i}^{c,d}(r)$ being carbon ${\rm p_z}$-orbitals located in the unit cell $i$ at sublattice A and B respectively. This LDOS of impurity resonances at $E_{\imp}=-0.1$\,eV for a single and scalar impurities are shown in figure \ref{fig:NallE01x6}. \cite{Illustr_U0}
The formation of virtual bound states (VBS) due to impurity scattering is clearly visible.

These VBS are a general feature of localized states hybridizing with a continuum of delocalized states. They have been observed in many systems ranging from elementary metals \cite{Gubanov1992} to d-wave superconductors \cite{balatsky:373}. Details of the real space image are however system specific. Here, the threefold ($D_{3h}$) symmetry of the VBS for a single impurity and twofold ($D_{2h}$) symmetry for a double impurity are direct consequences of the lattice symmetry. The ($D_{3h}$) symmetric single impurity state results in sixfold symmetric Fourier transformed scanning tunnelling spectra \cite{bena:2005}. Furthermore the peculiarities of the bandstructure of graphene manifest themselves in the near field characteristics of impurities:
A single impurity in sublattice A induces an impurity state mostly localized in sublattice B and vice versa due to the fact that $G^0_{11}(i,E)\ll G^0_{12}(i,E),G^0_{21}(i,E)$ for $E\rightarrow 0$, which can be attributed to the existence of two nonequivalent Dirac points as explained above.

\begin{figure}
\centering
\includegraphics[width=.98\linewidth]{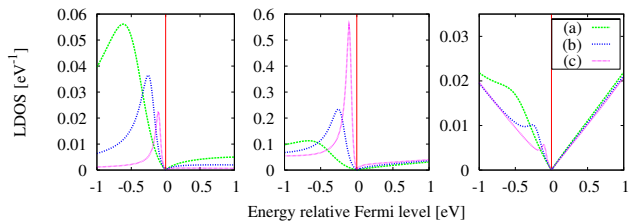}
\caption{\label{fig:Nsingle_impNN}(Color online) LDOS at the impurity site (left), NN (middle) and NNN sites (right) is shown for a single impurity with potential $U_0=10$\,eV (a), $20$\,eV (b) and $40$\,eV (c).}
\end{figure}

The site projected DOS $N(i,E)$ can be obtained from the full Green's function $N(i,E)=-1/\pi \Im G(i,i,E)$, where each of the diagonal matrix elements corresponds to one sublattice. For the single impurity and $U_0$ from $10$\,eV to $40$\,eV the LDOS at the impurity, nearest neighbour (NN) and next NN sites are shown in figure \ref{fig:Nsingle_impNN}. One sees that for vacancies with $10eV\lesssim U_0\lesssim 20$\,eV \cite{pogorelov-2006-}, but not for weaker potentials, an impurity resonance should be clearly observable within $1$\,eV around the Dirac point.

\begin{figure}
\centering
\includegraphics[width=.98\linewidth]{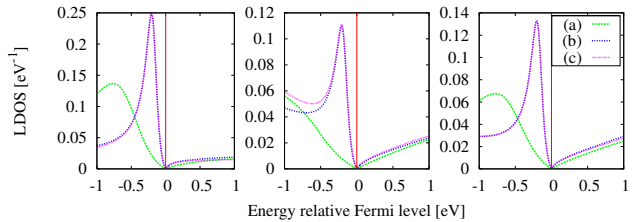}
\caption{\label{fig:Ndouble_impNN}(Color online) As figure \ref{fig:Nsingle_impNN} but for scalar double impurity with $U_0=4$\,eV (a) and $6$\,eV (c) as well as for double impurity with $U_0=4$\,eV and $U_1=-2$\,eV (b).}
\end{figure}

It further illustrates the localization of the impurity state on sublattice B, when the impurity is in sublattice A, as well as the reduction of LDOS at the impurity site for strong repulsive potential.
The double impurity respects pseudospin symmetry and is much more sensitive to weaker potentials as can be seen from figure \ref{fig:Ndouble_impNN}.
Clearly, $U_t=U_0-U_1$ is the most important parameter determining the shape of LDOS in the case of a double impurity: The results for ($U_0=4$\,eV, $U_1=-2$\,eV) and ($U_0=6$\,eV, $U_1=0$\,eV) are virtually indistinguishable at the impurity resonance but differ slightly below it. For large distances $r\gg\hbar v_{\rm f}/E_{\imp}$ from the impurity site we obtain for the changes in LDOS $\Delta N(r,E_{\imp})\propto 1/r$ in agreement with \cite{bena:2005, lin-2006} for all considered types of impurities. Note the contrast to a single hard-wall impuritiy, i.e. $U_0\rightarrow\infty$, with $1/r^2$ asymptotics of $\Delta N(r,E_{\imp})$ \cite{pereira2006}.

\begin{figure}
\centering
\includegraphics[width=.85\linewidth]{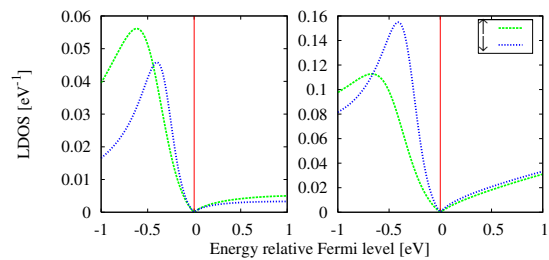}
\caption{\label{fig:Nsingle_mag}(Color online)  SP-LDOS at the impurity site (left) and a NN site (right) is shown for a single magnetic impurity with $U_0=12$\,eV and $J=2$\,eV.}
\end{figure}

If the impurities have a \textit{magnetic moment}, exchange scattering of the graphene ${\rm p_z}$-electrons and the spin $S$ localized at the impurity site will occur. As long as the exchange coupling $J$ does not exceed a critical value, Kondo screening of the spin $S$ by the band electrons can be neglected and the impurity spin acts as local magnetic field: The effective scattering potential is renormalized to $U_0\pm J$. The resulting change in spin-polarized (SP) LDOS in the vicinity of a single impurity is shown in figure \ref{fig:Nsingle_mag} for $U_0=12$\,eV and $J=2$\,eV.
The exchange splitting of the resonances in the two spin channels is approximately $0.15$\,eV. This type of exchange scattering also affects decay lengths and oscillation periods of the induced spin density variations and therefore provides a possible mechanism for long range exchange interactions.

\begin{figure}
\centering
\includegraphics[width=.85\linewidth]{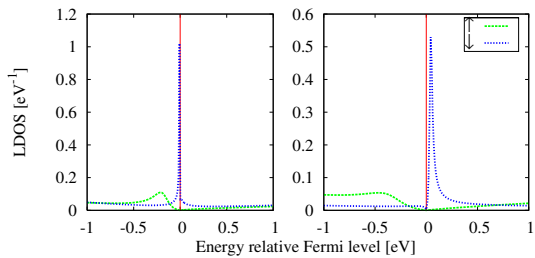}
\caption{\label{fig:Ndouble_mag}(Color online) SP-LDOS at NN site of a double impurity with $U_0=5$\,eV, $U_1=-2$\,eV and $J=1$\,eV (left) and $2$\,eV (right) are shown.}
\end{figure}

For double impurities the effect of exchange splitting is much more pronounced within a realistic parameter range: As figure \ref{fig:Ndouble_mag} shows, exchange scattering can produce strongly spin polarized impurity states. The impurity resonances of one spin channel can be pushed close to the Dirac point or the impurity levels are split even below and above it.
Depending on the type of impurities, the spin polarization of the impurity states can strongly depend on doping: In the example with $J=2$\,eV, the VBS above the Dirac point can be occupied by spin down electrons due to n-doping.

It was demonstrated recently \cite{Edwards_2006} that ferromagnetism
of sp electrons in narrow impurity bands can be characterized by
much higher Curie temperatures than those typical for traditional
dilute magnetic semiconductors. The reasons are a suppression of
T-matrix renormalization of the Stoner exchange parameter and a
high magnon stiffness constant. So the Curie temperature is
basically determined by single-particle Stoner excitations, in a
sharp contrast with d-electron magnets. Hence the impurity band
associated with the magnetic impurities considered in this letter
can be a promising candidate for facilitating high temperature
ferromagnetic order in graphene.

\textit{In conclusion},  we have calculated the LDOS of impurity
resonances in graphene from the near field to the regime of
asymptotic $1/r$ decay. The near field LDOS are directly observable
by STM and comparison of upcoming experiments with our predictions will
eludicate the nature of impurities in graphene. We also find that
impurity resonances in graphene are very different from the impurity
states observed in graphite because of the two sublattice structure
in graphene.

We showed further how spin polarized impurity states can result from exchange
scattering at magnetic impurities and their sensitivity to doping.
The resulting formation of spin polarized impurity bands may give
 rise to long range exchange interactions and magnetic order, that
  can be directly studied by spin-polarized STM.

We are grateful to M. Bode, A. Castro Neto, J. Fransson,
A. Geim, A. Kubetzka, K. Novoselov and J.X. Zhu for useful
discussions. This work has been supported by LDRD and DOE BES at
Los Alamos, FOM (Netherlands) and SFB 668. A.V.B is grateful to U. Hamburg and Wiesendanger group
for hospitality during the visit, when the ideas presented in this
work were conceived.
\bibliography{graphene.bib}

\begin{thebibliography}{33}
\bibitem[{Ill()}]{Illustr_U0}
\bibinfo{note}{$E_{\imp}=-0.1$\,eV requires $U_0=45$\,eV for a single impurity.
  As $U_0$ is only a fitting parameter for future experiments, the choice of
  $U_0=45$\,eV has been made to illustrate the resonance near Dirac point.}
\end{thebibliography}
\end{document}